\documentclass[10pt]{article}
\usepackage[left=2.5cm, right=2.5cm, top=2.5cm, bottom=2.5cm]{geometry}
\usepackage{setspace}
\usepackage{booktabs}
\usepackage{siunitx}
\usepackage[colorlinks,linkcolor=black,anchorcolor=black,citecolor=blue]{hyperref}
\usepackage{amsmath, amsfonts, amssymb} 
\usepackage{multirow}  
\usepackage{array}      
\usepackage{tablefootnote}
\usepackage[version=3]{mhchem} 
\usepackage{lastpage,fancyhdr,graphicx}
\usepackage{epstopdf}
\usepackage{color}
\usepackage{sidecap}
\usepackage{wrapfig}
\usepackage[pscoord]{eso-pic}
\usepackage[fulladjust]{marginnote}
\usepackage{subcaption}
\reversemarginpar

\begin{document}

\begin{flushleft}
{\huge\bfseries Multi-Dimensional Photon-Correlations Reveal Triexciton Features in Single Perovskite Quantum Dots}

\vspace{0.3cm}

{\large
Alex Hinkle\textsuperscript{1}, Chieh Tsao\textsuperscript{1,2}, 
Adam Duell\textsuperscript{1},
Hendrik Utzat\textsuperscript{1,2*}
}
\\
\bigskip
\textsuperscript{\textit{1}} Department of Chemistry, University of California, Berkeley, California 94720, USA
\\
\textsuperscript{\textit{2}} Materials Science Division, Lawrence Berkeley National Lab, Berkeley, California 94720, USA
\\
* \href{mailto:hutzat@berkeley.edu}{hutzat@berkeley.edu}

\end{flushleft}

% use {asbstract*} to suppress the copyright line. Copyright information will be added in production

\section*{Abstract} 
Lead-halide perovskite quantum dots (PQDs) are established quantum emitters with potential for entangled photon-pair generation via multiexciton cascades. However, the energetics and dynamics of many-body excitations remain poorly understood. Here, we perform time- and frequency-resolved photon-correlation spectroscopy of single CsPbBr\textsubscript{3} PQDs at low temperatures using a single-photon avalanche diode (SPAD) array detector. We report biexciton binding energies and assign their charged states, which undergo fast ($\mu$s) switching dynamics. Most notably, we identify a spectral feature blue-shifted from the exciton by $7.4 \pm 1.9$ meV as the bound triexciton and establish the order of its cascade emission. These results highlight the power of low-temperature, multidimensional photon-correlation spectroscopy for resolving complex many-body dynamics.

%As a proof of principle, we extract temperature-dependent emission lifetimes from HED, revealing dynamics consistent with thermal exchange between dark and bright exciton states.

%%%%%%%%%%%%%%%%%%%%%%%%%%  body  %%%%%%%%%%%%%%%%%%%%%%%%%%
\section{Introduction}
Colloidal lead–halide perovskite quantum dots (PQDs) have rapidly emerged as a versatile class of quantum light sources. Their appeal lies in a unique combination of solution-processable and compositionally tunable synthesis, high photoluminescence quantum yields, and remarkable optical stability compared to earlier colloidal systems \cite{kovalenko2015,protesescu2015,gintersedernano2021}. A particularly striking feature of PQDs is their large oscillator strength \cite{boehme2025single,becker2018bright}. At cryogenic temperatures, individual PQDs exhibit linewidths approaching the Fourier limit and long optical coherence times \cite{utzat2019coherent,raino2020}, raising prospects for their use in coherent quantum optics. More recently, two-photon Hong–Ou–Mandel interference experiments have directly demonstrated the feasibility of generating indistinguishable single photons from colloidal PQDs \cite{muller2023}, positioning these materials as a solution-processable alternative to epitaxial III–V quantum dots in the visible spectral range.\\
Multi-exciton dynamics in QDs has been studied to identify Auger quenching e.g. to maximize brightness in LEDs or identify exciton fission \cite{amgar2019higher, benning2025photon}. For coherently-emitting PQDs, possible entangled-photon generation from cascaded biexciton-exciton emission is another motivator. Cascaded emission was developed and refined in epitaxial III–V systems, where polarization entanglement emerges provided the intermediate bright-exciton doublet is degenerate or nearly degenerate \cite{benson2000,stevenson2006,akamatsu2020}. Extending such cascades to colloidal PQDs raises compelling opportunities for scalable quantum sources. However, realizing this potential requires precise control over multiexciton energetics, as well as their recombination order and rates, at the single-particle level. Compared to III–V nanostructures, PQDs may present a richer multi-exciton physics due to their strong coupling to lattice phonons and large polaron formation \cite{miyata2017,herz2018}. Emblematic of this are earlier debates on the binding energy of the biexciton, with different experimental methods identifying widely discrepant positive and negative values \cite{doi:10.1021/acsnano.6b03908, doi:10.1021/acs.jpclett.8b01029, doi:10.1021/acs.jpcc.7b00762}, which has since been resolved \cite{PhysRevB.111.155304, PhysRevMaterials.7.106002,lubin_resolving_2021,doi:10.1021/acs.jpclett.9b02015}. Lattice degrees of freedom cannot be ignored in anharmonic perovskite lattices. Phonon population may therefore substantially alter biexciton and higher-order exciton binding energies and, consequently, their role in entangled-photon cascades.\\
Photon-correlation spectroscopy has long served as a central tool for characterizing multiexciton emission in single quantum dots \cite{Lubin2022, dalgarno2008,utzat_review2025}. In II–VI nanocrystals such as CdSe, higher-order correlation measurements up to $g^{(3)}$, combined with emission color filtering, have been used to identify triexciton (TX) emission quantum yields and to place lower bounds on binding energies \cite{shulenberger2021,bawendi2000}. However, emission color filtering alone cannot provide detailed information about energy–time correlations along the emission cascade, nor can it unambiguously disentangle overlapping emission features. Recent studies have leveraged single-photon avalanche diode (SPAD) arrays in combination with spectrally dispersive elements. This approach enables time-tagged and photon-number-resolved coincidence measurements, thereby providing access to multiexciton spectra at the single-nanocrystal level \cite{Lubin2021,Lubin:19,scharf2025}. Despite its power, such higher-order and energy-resolved photon-correlation spectroscopy has not yet been widely adopted, largely due to the technical challenges of still-emerging SPAD array technology, cost, and the required extensive post-processing for artifact correction. Nonetheless, pioneering works have already extracted biexciton and triexciton binding energies, as well as characterized spectral diffusion of PQDs at room temperature \cite{Lubin2021,scharf2025, doi:10.1021/acs.nanolett.4c03096}. One of remarkable findings to emerge through this method is the identification of red-shifted triexciton emission in large CsPbBr$_3$ QDs at room temperature \cite{doi:10.1021/acs.nanolett.4c03096}. This behavior runs counter to the expectation from Pauli repulsion, which would require one electron–hole pair to occupy higher-lying excitonic states and could thus lead to a blue-shifted emission. From this perspective, low-temperature studies of multiexciton physics in PQDs can provide complementary insight under conditions of reduced phonon population and minimized lattice disorder. By suppressing thermal fluctuations, such measurements can isolate the intrinsic binding from Coulomb and exchange interactions in multiexcitons.\\
Here, we extend heralded single-dot spectroscopy to cryogenic temperatures, implemented with a SPAD array coupled to a spectrometer. We further conduct time- and frequency-resolved $g^{(2)}$ analysis that reveals the time-ordering of multi-exciton cascades. This approach allows the unambiguous identification of biexciton and triexciton emission cascades and binding energies, as well as multi-timescale charging kinetics, and spectral diffusion without the spectral congestion from phonon broadening.

\section{Results}
\begin{figure}[h!]
    \centering\includegraphics[width=16cm]{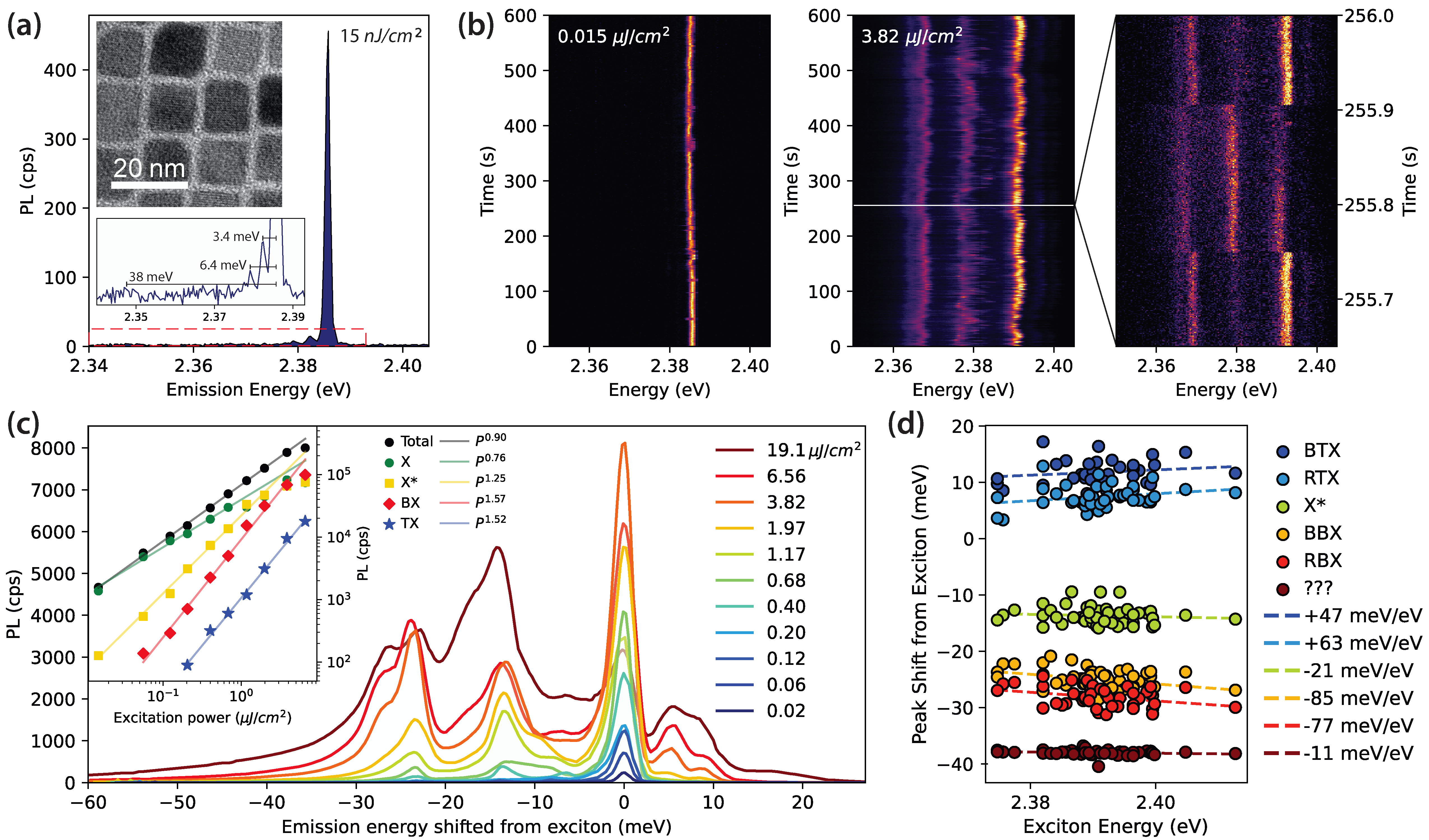}
    \caption{\textbf{Single PQD spectral characteristics at 6K.}(a) shows a representative TEM and single-frame emission spectrum displaying resolution-limited ZPL manifolds with three distinct lower-energy side bands. (b) Spectral time series taken with the SPAD array at two different fluences. Several additional peaks emerge at higher fluence and spectral diffusion over several meV becomes prominent. Fast (single-ms) spectral shifting dynamics are observable with the 1 ms frame rate. (c) Single PQD spectra at different excitation fluences and intensity dependence of different peaks. The tallest peak is assigned to the exciton fine-structure manifold. (d) Peak energies as a function of the exciton energy for 50 dots. Fit lines are calculated using a simple linear regression.}
    \label{fig1}
\end{figure}

We synthesized CsPbBr\textsubscript{3} nanocrystals according to Ref. \cite{gintersedernano2021}. PQDs with C3-bridged quaternary ammonium bromide ligands exhibit improved benchtop stability, as well as enhanced stability and reduced spectral jitter under high excitation fluence. The PQDs had an average edge length of 15.46 $\pm$ 1.76 nm. At room temperature the particles have an ensemble emission peak at 517.5 nm with a FWHM of 21 nm. The PLQY of was 58\% under 450nm excitation. Our PQDs’ high stability is exemplified in Fig. \ref{fig1}, which shows representative single-particle characterization under non-resonant 484 nm pulsed excitation (sub-ps pulse width, 80 MHz) at 6 K. Low-excitation-power emission spectra Fig. \ref{fig1}a consistently displayed narrow (1.2 meV FWHM) lines limited by the resolution of our SPAD-spectrometer (0.3 meV/pixel). Earlier work has identified two or three zero-field exciton fine-structure states with splittings of 0.2–2 meV for tetragonal and orthorhombic single-particle polymorphs \cite{becker2018bright, doi:10.1021/acs.nanolett.7b00064} PQDs with C3-bridged diquat ligands are known to display comparatively smaller fine-structure splittings of up to hundreds of $\mu$eV, too small to be resolvable with most spectrometers \cite{ginterseder_lead_2023}. Fig. \ref{fig1}a further reveals lower-energy phonon sidebands at $\approx 3.4$ and $\approx 6.4$ meV, previously assigned to two LO phonon modes \cite{cho2022exciton, amara2023spectral}. The low Huang–Rhys parameters ($<0.05$) indicate minimal exciton–phonon coupling to these modes. Following the agreement of the two clear phonon modes with previous literature, we might expect an additional $-19$ meV LO mode \cite{cho2022exciton, amara2023spectral}. While there may be a slight signal in that region, it is difficult to identify above the baseline noise. Interestingly, we do observe a spectral feature $-38.0 \pm 0.5$ meV from the exciton manifold ($E_X$). This feature is remarkably consistent in both occurrence and shift from the exciton. It appears at very low fluences, and while present at higher fluences, it is sometimes obscured due to overlap with neighboring peaks.  At roughly double the energy and showing little size dependence, the $-38.0$ meV peak is potentially the second harmonic of the $19$ meV LO phonon. 

The use of high-stability C3-bridged PQDs is critical for our study of many-body interactions under high fluence. Fig. \ref{fig1}b shows spectral time series at different excitation fluences. Spectral diffusion with meV-scale amplitude is observed in all cases and becomes more pronounced with increasing excitation fluence, accompanied by the emergence of additional spectral features. Based on previously published binding energies \cite{https://doi.org/10.1002/adma.202208354}, the peaks $-13.7 \pm 1.5$ meV and $-25.0\pm 1.8$ meV of $E_X$ can be preliminarily assigned as the trion (X*) and biexciton (BX), respectively. Trion formation in QDs is a discrete, stochastic, multi-timescale process. The high timing resolution and absence of readout noise of the SPAD array enable spectral analysis with high ($>$1000 per second) frame rates, as shown. This allows us to capture sub-second spectral jumps, as seen in Fig. \ref{fig1}b.

\indent Further systematic variation of the excitation fluence revealed the super-linear intensity increase of the $25.0$ meV red-shifted peak, as seen in Fig. \ref{fig1}c (exponent $p=1.57$). This confirms the peak assignment to the BX. Especially notable at high excitation powers, the biexciton appears to split into a main peak and a low-energy shoulder ($-28.1 \pm 1.8$ meV from $E_X$), referred to here as the blue biexciton (BBX) and red biexciton (RBX), respectively. We will explain the deviation from $p=2.0$ expected for the BX further below.

\indent The $13.7$ meV red-shifted peak assigned to the trion increases remarkably consistently with the excitation power ($p=1.25$). This is unexpected due to its stochastic population often occurring on second- or even slower-timescales. The observed power-law behavior therefore hints at a light-driven trion population with sufficiently fast rates to fully sample the exciton–trion equilibrium during our data acquisition time (1000 s). At higher excitation fluences of $>1.17 \mu J/cm^{2}$, two peaks, $7.4 \pm 1.9$ meV and $11.8 \pm 2.1$ meV blue-shifted from $E_X$, become noticeable. The presence of a power-dependent blue-shifted peak has been published before but not explicitly commented upon \cite{https://doi.org/10.1002/adma.202208354}. The sum intensity of these peaks also increases super-linearly with $p=1.52$. As will be expanded on later, we assign these blue-shifted peaks to the triexciton despite their deviation in $p$ from $p=3.0$. We will refer to the two peaks as the blue triexciton (BTX) and red triexciton (RTX). We note that the sum intensity over all peaks remained in the linear regime for powers $<6.56 \mu J/cm^{2}$ and that any peak broadening below that power was reversible.

\indent Fig. \ref{fig1}d also displays the relative peak separations from $E_X$ for the X* and red and blue BX and TX for a sample of 50 single dots. Increasing exciton energy blue-shifts the TX and red-shifts the BX, but more definite quantification is hampered by comparatively large particle-to-particle variation versus the smaller range of surveyed $E_X$.

\begin{figure}[h!]
    \centering\includegraphics[width=16cm]{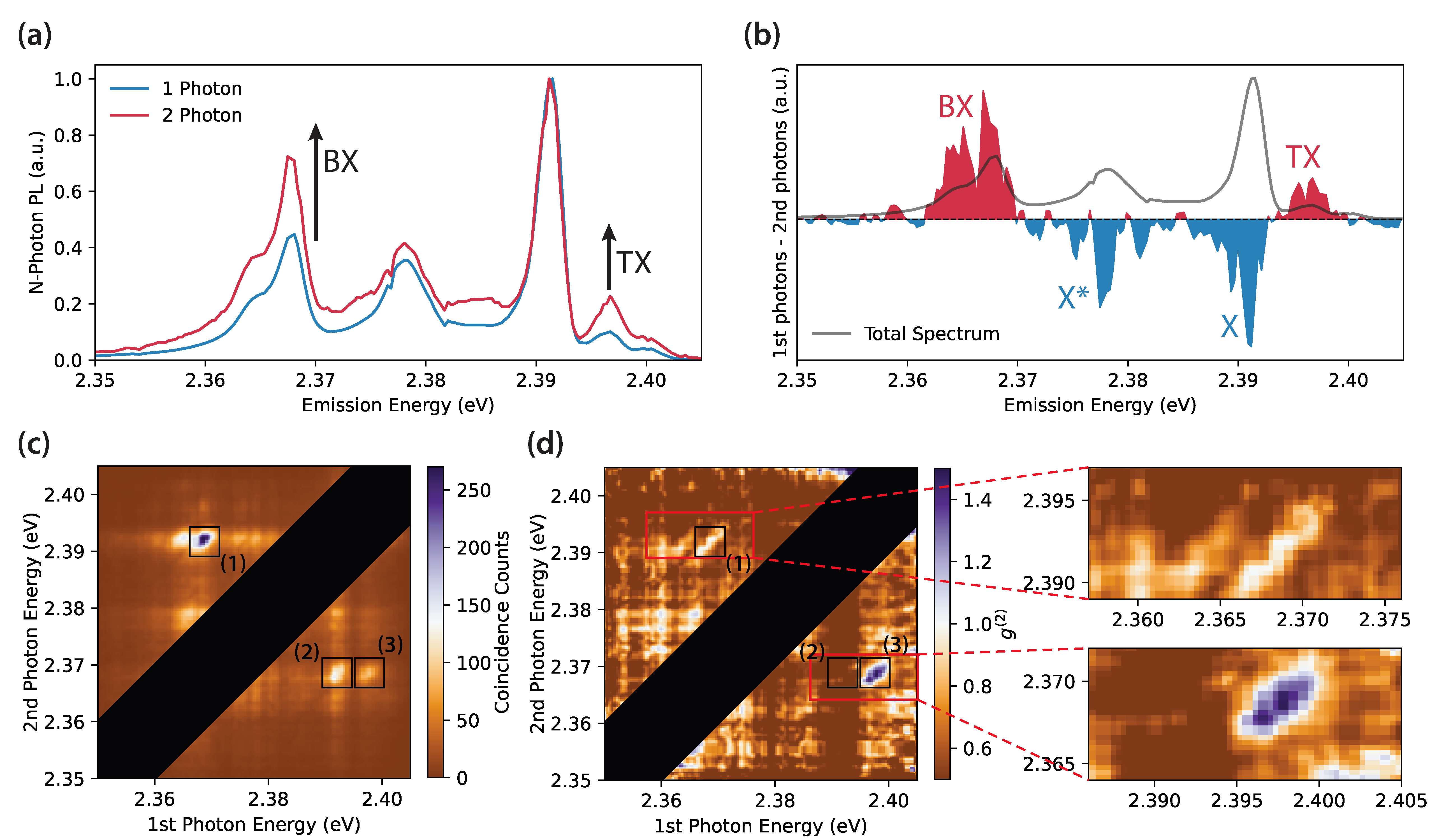}
    \caption{\textbf{Frequency-resolved $g{(2)}$ analysis around $\tau=0$.} (a) Emission spectra for single- and two-photon emission events normalized to the X peak. (b) Time-ordered difference spectrum between first and second photons in two-photon events, revealing the spectral position of second photons offset by -26.9/-23.4 meV (red/blue BX) and +5.2 meV, assigned to the TX. (c) Un-normalized frequency-resolved correlation maps between first and second photons indicating several cross-peaks. We indicate relevant transitions with black squares: BX/X  (Feature 1), X/BX (Feature 2), and TX/BX (Feature 3). (d) Normalized frequency-resolved g(2) function. Diagonal features between the TX/BX become more visible. The zoomed in regions around the BX/X (Feature 1) and TX/BX transitions (Feature 3) reveal a diagonal shape. Data along the diagonal are obscured by detector artifacts and dark counts and removed for clarity.}
    \label{fig2}
\end{figure}

\indent Our analysis thus far has revealed blue-shifted spectral features with super-linear intensity increase, but suppressed power-law coefficients, which is insufficient to justify the assignment to a bound TX. We therefore turn to spectroSPAD for photon-number-resolved spectroscopy. After pioneering work at room temperature\cite{Lubin2021, doi:10.1021/acs.nanolett.4c03096}, our application to delineate cascade pathways at cryogenic temperatures, where phonon broadening is minimized, significantly expands the scope of this powerful method.
\indent Fig. \ref{fig2}a shows the integrated spectra for all laser pulses followed by one and two photons. We refer to these as the “one-photon spectrum” and “two-photon spectrum,” respectively. The BX and TX peaks are more pronounced in the two-photon spectrum, suggesting preferential emission at these wavelengths for two-photon detection events. A more convincing representation further takes the time-ordering of the detected two-photon events into account. Fig. \ref{fig2}b shows a time-ordered difference spectrum between first and second photons out of two-photon events. Even though the SPAD array’s temporal jitter ($\approx 100$ ps) is on the order of the X lifetime (100-200 ps), preventing definite arrival-time sorting at these timescales, time-ordered difference spectra reveal a clear clustering of first photons around the BX and TX peaks. This data can only be explained if the TX peak is part of a cascade emitting at least two photons.

\indent To confirm our TX assignment, we conducted two-dimensional frequency-resolved $g^{(2)}$ analysis, as seen in Fig. \ref{fig2}c, which displays the energy of the first and second photons of two-photon coincidences. To the best of our knowledge, this type of analysis has not been conducted yet, likely due to the absence of narrow features at room temperature and the substantive artifact correction not being readily available \cite{Lubin:19}. Cross-peaks in these coincidence maps show two-photon cascades with different photon energies. The prominent Feature (1) corresponds to the biexciton–exciton cascade. Feature (2) presents as a less intense cross-peak between the X and BX, and Feature (3) as a cross-peak between the TX and BX. While Feature (1) is consistent with the BX-to-X cascade, an appreciable Feature (2) is not. Therefore, Feature (3) in this representation is insufficient to confirm the TX assignment. We show the same dataset normalized by the expected number of accidental coincidences based on the average count rates at a given wavelength in Fig. \ref{fig2}d. The color bar therefore indicates raw coincidences in (c) and the normalized $g^{(2)}$ in (d). In this representation, Feature (3) is larger in amplitude than Feature (1), indicating a higher prevalence of TX-to-BX cascade events over BX-to-X events. Most importantly, Feature (2) is not only suppressed, but a $g^{(2)}<0.7$ indicates anti-bunching in the cross-correlation between X-to-BX emission events. We therefore conclude that the TX-to-BX cross-peak is a significant feature and that, when a TX is excited statistically, P-state emission preceding BX S-state emission is the dominant cascade pathway.

\indent The zoomed-in regions in Fig. \ref{fig2}d of Features (1) and (3) further emphasize the diagonal shape of the cross-peaks. Non-circular features in our analysis indicate persistent energy correlations between cascading photons. Right-diagonal shapes, as observed here, indicate positive correlations between the cascading photons, for example caused by correlated spectral diffusion between the two cascaded states. We suggest that this analysis of the joint spectral density is well-suited for the identification of energy anti-correlations, as also expected in entangled pairs, which we do not observe here.

\indent Altogether, our analysis suggests a blue-shifted TX with fast emission to the BX. However, the assignment remains inconsistent with the power-law coefficients of $<3$ and $<2$ for the TX and BX, respectively. To further explain the red and blue shoulders observed in the BX and TX (RBX, BBX, and BTX, RTX), we hypothesize that an excitation-power-dependent switching rate from neutral to charged excitons can reduce the power-law coefficients and indicate the charging state of the shoulders through their intensity (anti-)correlation with the X and X* states.

\begin{figure}[h!]
    \centering\includegraphics[width=16cm]{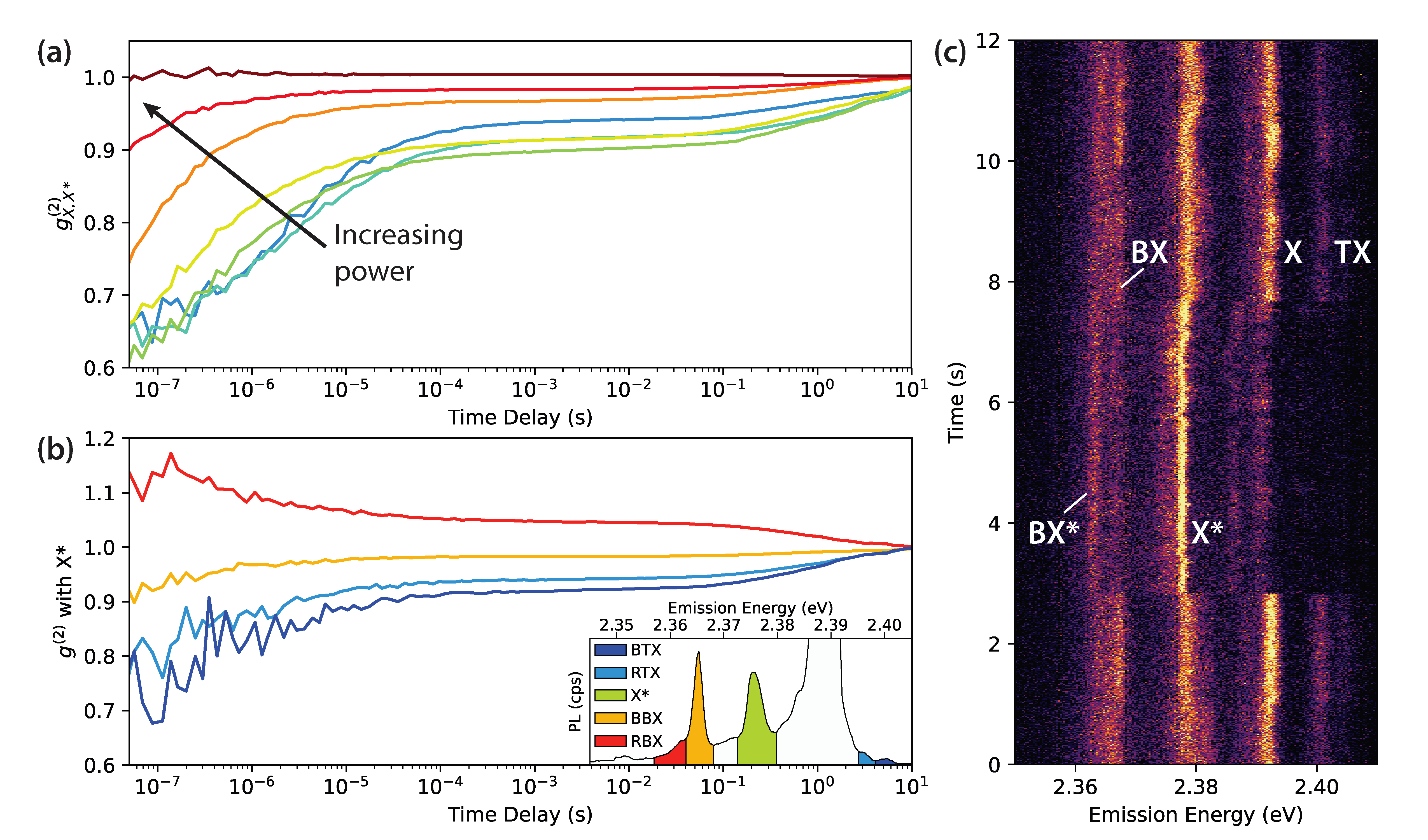}
    \caption{\textbf{Frequency-resolved log-space intensity correlation between different spectral features.} (a) Excitation power dependence of the $g^{(2)}_{X,X*}$ between X and X* indicating antibunching below tens of microseconds to hundreds of nanoseconds. Power series conducted from $0.4$ to $19.1 \mu J/cm^{2}$. (b) Cross-correlations between select spectral features and X* at $2 \mu J/cm^{2}$. Positive correlation indicates the charged character of the species, and negative correlation a neutral character. (c) Spectral time series around a distinctly charged state, mirroring the (anti-)correlations in (b). }
    \label{fig3}
\end{figure}

\indent Fig. \ref{fig3}a shows the normalized log-spaced intensity cross-correlation between the X and X*, with two inflection points: one on the seconds timescale and one between hundreds of nanoseconds and tens of microseconds, both dependent on excitation intensity. Early-$\tau$ anti-bunching suggests the lack of simultaneous trion and exciton emission, true by definition. Similarly, we may expect bunching or anti-bunching between the RBX and BBX when correlated with the trion. Fig. \ref{fig3}b shows these cross-correlations for select spectral features and the X* at constant excitation fluence (1.97 $\mu J/cm^{2}$). The RBX shows clear correlation with the X*, while the BBX shows anti-correlation, confirming our assignment of the RBX as the charged BX and the BBX as the neutral species. This interpretation is supported by the spectral time series in Fig. \ref{fig3}c, where a long event of persistent trion emission coincides with increased RBX emission and suppression of BBX.

\indent The same analysis is less conclusive for the TX spectral region. As seen in Fig. \ref{fig3}b (inset), the TX overlaps with the tail of the X, complicating correlation analysis. Both RTX and BTX exhibit anti-correlation with the trion, suggesting that neither are charged species. However, the marginal differences between their correlations with X* prevent confident assignment of charging states. Moreover, the TX region appears highly quenched during trion emission events (Fig. \ref{fig3}c), suggesting that charging of the TX activates Auger recombination. Two smaller peaks slightly red-shifted from $E_X$ also appear during charging, as seen in Fig. \ref{fig3}c. While the timing of their appearance and similar spectral shape might suggest charged triexcitons, this is not supported by other evidence: (a) similar peaks occur during charging events even at low fluences, and (b) no distinct correlation is seen in the frequency-resolved $g^{(2)}$ spectra (Fig. \ref{fig2}a-b) for these peaks being followed by X* emission.

\indent Finally, we note that the apparent strength of the (anti-)correlations is greater for the TX than for the RBX and BBX. This likely arises from two factors: (1) the BX peaks overlap more strongly with each other and with the trion tail, which reduces the level of correlation, and (2) residual emission from the charged or uncharged BX is observed in Fig. \ref{fig3}c even when the exciton or trion dominates. It is not clear whether this is due to fast switching between charged and uncharged states, or if there are other currently unidentified peaks coincidentally overlapping with these regions (e.g. BX phonon branches, S-state TX emission).

%\indent Cascade emission in these spectra appears as cross-peaks... DISCUSS. One-, two-, and three-photon spectra cannot inform on the net charging of a given emissive species. To further assign the shoulders of the biexciton and triexciton peaks to their respective charged species, we turn to longer-timescale correlation analysis. We observed the seamless onset of trion emission at intermediate excitation fluences, in most cases without resolvable switching dynamics between X and X*, indicating switching rates faster than the frame rate of spectral acquisition (see Fig. 1C). This is confirmed in Fig. 3A, where we show the intensity cross-correlation between sets of peaks around the X and X* emission, revealing the clear anti-correlation expected from discrete switching events.

%Time-and frequency-ordered $g(2)$ maps also enable the ordering of the full TX emission cascade. Here, proper normalization of the frequency-resolved $g(2)$ is required to obtain information on time-ordering.

\section{Discussion}
\indent The value of this work lies in the detailed analysis of many-body exciton dynamics and the first demonstration of cryogenic time- and frequency-ordered $g^{(2)}$-analysis, which has broad utility in various excitonic materials. We first discuss our results in the context of previous work on multiexcitons in perovskite quantum dots. Most prior studies were performed at room temperature, identifying fast Auger recombination as a function of size and halide composition \cite{lubin_resolving_2021, utzat2017probing, cho2024size}, sometimes with $g^{(2)}(\tau=0)<0.05$ \cite{zhu2022room}, indicating high-purity single-photon emission. Emission speedup at low temperatures is known to reduce overall multiexciton quenching \cite{hu2016slow}, rendering multiexciton species comparatively bright in contrast to III–V and II–VI quantum dots rendering them directly observable \cite{doi:10.1021/acs.nanolett.7b00064, https://doi.org/10.1002/adma.202208354}. The charged biexciton has also been reported through its distinct spectral position \cite{yin2017bright}.Blue-shifted spectral features have been observed in several studies, but they have never been definitively assigned or discussed in detail. For example, \textit{Zhu et al.} observed a +8 meV blue-shifted peak in CsPbX$_3$ of 23nm edge length with appreciable simultaneous BX emission\cite{https://doi.org/10.1002/adma.202208354}. Other studies do not record the blue-shifted region \cite{hu2016slow}.

\indent Our observations support the existence of a bound triexciton at low temperatures, in which Pauli repulsion enforces higher-energy $p$-exciton occupation with one electron and one hole. The difference in the average energies of the lower-energy TX ($+7$ meV) and the neutral BX ($-25$ meV) suggests an $S$–$P$ exciton energy difference of at least $+32$ meV. This estimated splitting is remarkably consistent with reported splittings in absorption spectra \cite{boehme2025single}. Our reported values for the TX energy are significantly smaller than those for CdSe QDs ($>100$ meV) \cite{shulenberger2021resolving, bonati2005spectral}. While we cannot fully delineate the effects of confinement, screening, polaron formation, and effective mass differences, the smaller TX energy is in line with the smaller exciton and biexciton binding energies observed across perovskite nanocrystals at room temperature \cite{carwithen2023confinement}.

\indent Our identification of a bound triexciton with appreciable, though as yet unquantified, emission quantum yield presents an interesting opportunity for further studies of many-body dynamics at low temperatures. The most remarkable effects in PQDs stem from coherent exciton motion \cite{boehme2025single}, phase locking, and the formation of collective phases (e.g., in superfluorescence) \cite{raino2020}. Our high-stability PQDs hosting emissive TX states can therefore extend these studies to excitons of different symmetry. 

\indent Finally, we note the difference between our results and earlier reports of red-shifted triexciton emission obtained at room temperature. These differences may be explained by the unique carrier–lattice interactions in perovskites, which have been invoked in polaron-assisted carrier shielding \cite{zhu2016screening} and have also been discussed in the context of biexciton binding energies \cite{park2023biexcitons}.

\indent In conclusion, we have introduced low-temperature spectroSPAD to compile previously unreported two-dimensional time- and frequency-resolved $g^{(2)}$ correlation maps. This new approach is uniquely suited to studying emission cascades in a variety of electronic nanomaterials. 

\section{Methods}
The synthetic procedures herein were developed from previous reports by Ginterseder \cite{ginterseder_lead_2023}.

\indent \textit{Chemicals:}\
Lead (II) acetate trihydrate (99.999\%), cesium carbonate ($\geq$99.9\%), triphenylphosphine dibromide (TPPBr\textsubscript{2}, 96\%), trioctylphosphine (TOP, 97\%), polystyrene (average MW $\sim$280,000), 1-octadecene (ODE, Technical Grade 90\%), toluene (anhydrous, 99.8\%), ethyl acetate (anhydrous, 99.8\%), acetonitrile (ACN, anhydrous, 99.8\%), and diethyl ether (anhydrous, $\geq$99.0\%) were purchased from Sigma-Aldrich. Didodecylmethylamine ($>$85.0\%), 1,3-dibromopropane ($>$98.0\%), and acetone (HPLC grade) were purchased from Tokyo Chemical Industry. Oleic acid (Technical grade 90\%) was purchased from Thermo Fisher Scientific. Chloroform-D (D, 99.8\%) was purchased from Cambridge Isotope Laboratories Inc. All chemicals were used as received without further purification.\

\indent \textit{C3-2C12 Ligand Synthesis:}\
A 15 mL three-necked flask was connected to a Schlenk line with a reflux condenser, rubber septa, immersion thermometer, and a magnetic stir bar before purging with N\textsubscript{2}. All glass joints were greased, wrapped in Teflon tape, and clipped. Didodecylmethylamine (8 mmol, 4 eq), 1,3-dibromopropane (2 mmol, 1 eq), and anhydrous ACN (5 mL) were added. The solution was refluxed under N\textsubscript{2} while stirring overnight. The solution was cooled to room temperature and diethyl ether (~30 mL) was added to precipitate the product. The mixture was centrifuged at 9 krpm for 1 min, and the supernatant was discarded. Fifteen mL of diethyl ether were added and the mixture was sonicated for 15 min. The mixture was again centrifuged at 9 krpm for 1 min and the supernatant discarded. This final washing step was repeated an additional two times to yield a bright white solid. The solid product was dried under vacuum overnight and bottled.

\textit{Cesium Oleate Synthesis (CsOl):}\
A 100 mL three-necked flask was connected to a Schlenk line and equipped with a glass neck, immersion thermometer, rubber septa, and a magnetic stir bar. Cesium carbonate (Cs\textsubscript{2}CO\textsubscript{3}, 1.63 g, 5 mmol, 2 eq) and oleic acid (30 mL, 96 mmol, 9.6 eq) were added to the flask and degassed under vacuum for 1 hr at room temperature. The temperature was raised to \SI{50}{\celsius} and held for 1 hr, then to \SI{110}{\celsius} and held for 3 hr. The final pressure was 50 mtorr. The crude solution was cooled to room temperature and transferred to a 50 mL centrifuge tube. The tube was filled with ACN, shaken, and centrifuged at 10 krpm for 5 min. Two layers formed and the upper ACN layer was pipetted off and discarded. Twenty mL of ACN was added to the remaining yellowish oil layer, shaken, and centrifuged at 10 krpm for 5 min; the upper ACN layer was discarded. This step was repeated twice more. Forty mL acetone was added to precipitate the product from the oil layer. The mixture was sonicated for 3 min and centrifuged at 11 krpm for 1 min, and the supernatant discarded. Fifteen mL acetone was added, sonicated for 3 min, then centrifuged at 11 krpm for 1 min. This step was repeated twice more. The resulting white powder was dried under vacuum overnight and stored under N\textsubscript{2}. If the product does not precipitate when exposed to acetone, the ACN wash procedure can be repeated until precipitation occurs.

\textit{Lead Oleate Synthesis (PbOl\textsubscript{2}):}\
A three-necked flask was connected to a Schlenk line and equipped with a glass neck, immersion thermometer, and rubber septa. Lead acetate trihydrate (Pb(OAc)\textsubscript{2}·3H\textsubscript{2}O, 4.6 g, 12 mmol, 1 eq) and oleic acid (30 mL, 96 mmol, 8 eq) were added to the flask and degassed for 1 hr at room temperature. The solution was heated to \SI{50}{\celsius} and held for 1 hr, then raised to \SI{70}{\celsius} for 50 min. The temperature was then raised to \SI{110}{\celsius} and held for 10 min before cooling to room temperature. The final vacuum pressure was 53 mtorr. The crude solution was split into two 50 mL centrifuge tubes and 30 mL acetone was added to precipitate the product. The mixture was centrifuged at 10 krpm for 2 min and the supernatant discarded. The crude pellets were combined and further processed. Fifteen mL of acetone was added to the pellet and sonicated for 15 min before centrifuging at 10 krpm for 2 min, and the supernatant discarded. This step was repeated twice more. The resulting bright white powder was dried under vacuum overnight and stored under N\textsubscript{2}.

\textit{Triphenylphosphine Dibromide Solution (TPPBr\textsubscript{2} Solution):}\
Inside an N\textsubscript{2} glove box, triphenylphosphine dibromide (500 mg, 1.2 mmol), trioctylphosphine (1.1 mL), toluene (2.2 mL), and a magnetic stir bar were added to a 6 mL septum-cap vial. The vial was sealed and heated to \SI{80}{\celsius} while stirring until all solids were dissolved. The solution was prepared just prior to the nanoparticle synthesis described below and was used while still warm.

\textit{C3-2C12 PQD Synthesis:}\
A 25 mL three-necked flask attached to a Schlenk line was equipped with a glass neck, rubber septa, an immersion thermometer, and a magnetic stir bar. Twelve mL of 1-octadecene (ODE) was added to the flask and degassed under vacuum ($<$55 mtorr) for 30 min at \SI{110}{\celsius}. The solution was filled with N\textsubscript{2} and cooled to room temperature. To the flask were added PbOl\textsubscript{2} (385 mg, 0.5 mmol), CsOl (132 mg, 0.32 mmol), and C3-2C12 ligand (48 mg, 0.05 mmol). The mixture was held under vacuum at \SI{50}{\celsius} for 30 min, then heated to \SI{100}{\celsius} and held for 45 min. The vacuum was switched to N\textsubscript{2} and the solution was heated to \SI{170}{\celsius}; 1.5 mL of TPPBr\textsubscript{2} solution was injected, and the reaction vessel was immediately quenched in an ice bath and cooled to room temperature for washing.

PQD Washing: The crude mixture was centrifuged at 10 krpm for 10 min, and the supernatant was discarded. Ten mL of toluene was added to the pellet and agitated to disperse the particles. Twenty mL (2× volume) of ethyl acetate was added to precipitate the particles, which were centrifuged at 10 krpm for 3 min, and the supernatant discarded. Five mL of toluene was added to disperse the pellet, then 10 mL ethyl acetate was added and centrifuged at 10 krpm for 3 min, and the supernatant was discarded. This step was repeated once more. The resulting pellet was dispersed in 5 mL toluene and centrifuged at 4 krpm for 1 min, and the supernatant was retained. The final solution was filtered using a 0.22 $\mu$m syringe filter, purged with N\textsubscript{2}, and stored in an \ce{N2} glove box.\

\textit{Characterization:}\

\indent \textit{Optical:}\
Absorption and photoluminescence quantum yield (PLQY) were measured using a Shimadzu UV-3600i Plus UV-Vis spectrometer and a Horiba Scientific FluoroMax SpectroFluorometer equipped with an integrating sphere, respectively. All measurements were made in anhydrous toluene. Prior to measurement, PQDs were diluted to an OD of less than 0.1. A white-capped 1 cm{$^2$} quartz Hellma cuvette was used for both measurements. Absolute PLQY ($\Phi$) was measured using 450 nm excitation and calculated using the following formula:

\begin{equation}
\Phi = \frac{E_C - E_A}{L_A - L_C}
\label{eq:plqy}
\end{equation}

where $L_A$ is the integrated excitation light from a blank measurement, $L_C$ is the integrated excitation light with the sample, $E_A$ is the stray emission from the blank, and $E_C$ is the total emission from the sample.

\indent \textit{Nuclear Magnetic Resonance (NMR):}\
\textsuperscript{1}H and \textsuperscript{13}C spectra were collected with a NEO-500 MHz NMR operating at 500 and 126 MHz, respectively. The instrument was equipped with a Bruker Avance IV NEO console and a 5 mm \textsuperscript{1}H/BB iProbe. C3-2C12 samples were dissolved in deuterated chloroform for analysis.

\indent \textit{TEM:}\
TEM images were taken on an FEI Tecnai 12 TEM at 120 kV. Undiluted particles in toluene were drop-cast onto a 400-mesh carbon-coated copper TEM grid (Electron Microscopy Sciences).

\indent Single-Particle Sample Preparation: \
In a nitrogen glovebox, the stock solution of PQDs was initially diluted in toluene until the absorbance of the first exciton peak was 0.02. Then, 100 $\mu$L of the diluted stock was diluted into 900 $\mu$L of a 5\% w/w solution of polystyrene in toluene, and 100 $\mu$L of that was diluted into another 900 $\mu$L of toluene. Finally, 50 $\mu$L of the prepared solution was spin-coated on optical-grade quartz for 30 s at 3000 rpm + 60 s at 6000 rpm. The samples were made less than a day before measurement and sealed in a plastic bag before removal from the glovebox to prevent unnecessary air exposure.

\indent Single-PQD Characterization:\
Samples were cooled to 6 K using a closed-cycle liquid helium cryostat (CryoAdvance® 50, Montana Instruments). Measurements were performed using a home-built confocal scanning fluorescence microscope. The particles were excited with a pulsed laser (484 nm, 80 MHz, 400 ps pulse width, FemtoFiber® Pro TVis, Toptica) that was spectrally cleaned with 480-10 nm and 488-10 nm bandpass filters (Thorlabs). For power series, variable neutral density filters were used to attenuate the power (Thorlabs). The emission was directed through a 50 $\mu$m pinhole and spectrally filtered using a 500 nm longpass and 650 nm shortpass (Thorlabs). The emission was then sent into a spectrograph (Acton SP-2300i, 1200 g/mm, Princeton Instruments) with the SPAD-Array (SPAD Lambda, Pi Imaging) at the output.

\indent SPAD Array Corrections:\
Dark counts from the SPAD-Array were found to contribute significantly to cross-talk (false detection on a second pixel) and after-pulsing. As such, pixels with over 5,000 dark counts per second were electronically disabled (38/320 pixels). For correlation and number-resolved measurements, the limit was reduced to 1,000 dark counts per second (54/320 pixels). The emission spectrum was carefully placed to minimize overlap with the disabled pixels as much as possible.

Cross-talk (occurring at $\tau \lesssim 0.3$ ns) was used to sync the coarse time of the pixels, and a pulsed laser diffused across the array was used to sync the fine time of the pixels. An electronic time gate of 2 ns was applied during data collection, along with an additional post-processing time gate to remove detection events that passed the electronic gate.

While there are published methods to remove cross-talk artifacts \cite{Lubin:19}, these approaches were found to be insufficient for correcting the output of this device, as cross-talk rates vary drastically from pixel to pixel (0.3\% to 2\% for nearest neighbors) and can extend across large distances. There is a small flat probability ($\sim$0.003\%) of cross-talk occurring across the entire length of the array, contributing to the broad background seen in the N-photon spectra and frequency-resolved $g^{(2)}$. Within 10 pixels, the cross-talk rate increases exponentially. As such, for all photon-number-resolved data analysis methods, a condition was imposed that the $n$ photons must all be at least 15 pixels apart to be counted. This unfortunately precludes photon-number-resolved analysis of the same or neighboring peaks.

The dead time per pixel is $\sim$16–20 ns, after which after-pulsing spikes and then quickly falls. As such, the log-scale correlation functions were cut at 50 ns, after which the after-pulsing contribution is negligible compared to the real signal.

\section*{Acknowledgments}
This work was financially supported by the College of Chemistry at the University of California, Berkeley and the US Department of Energy (DE-AC02-05CH11231). The authors thank the staff at the University of California Berkeley Electron Microscope Laboratory for advice and assistance in electron microscopy data collection. We thank Drs. Hasan Celik, Raynald Giovine, and the Pines Magnetic Resonance Center’s Core NMR Facility (PMRC Core) for spectroscopic assistance, and Matthias Ginterseder for synthesis guidance. The NMR instruments used in this work were supported by the PMRC Core.

\section*{Supplemental document}
The data supporting the findings of this study are available within the article and its Supplementary Information.

\bibliographystyle{unsrt}
\bibliography{refs}
\newpage

\section*{Supplementary materials}

\renewcommand{\thesection}{S\arabic{section}}
\renewcommand{\thesubsection}{\thesection.\arabic{subsection}}
\renewcommand{\thesubsubsection}{\thesubsection.\arabic{subsubsection}}
\renewcommand{\theequation}{S\arabic{equation}}
\renewcommand{\thetable}{S\arabic{table}}
\renewcommand{\thefigure}{S\arabic{figure}}

% Reset counters
\setcounter{section}{0} 
\setcounter{subsection}{0}
\setcounter{equation}{0}
\setcounter{table}{0}
\setcounter{figure}{0}

\begin{figure}[h!]
    \centering
    \begin{subfigure}[t]{0.49\textwidth}
        \centering
        \includegraphics[width=\textwidth]{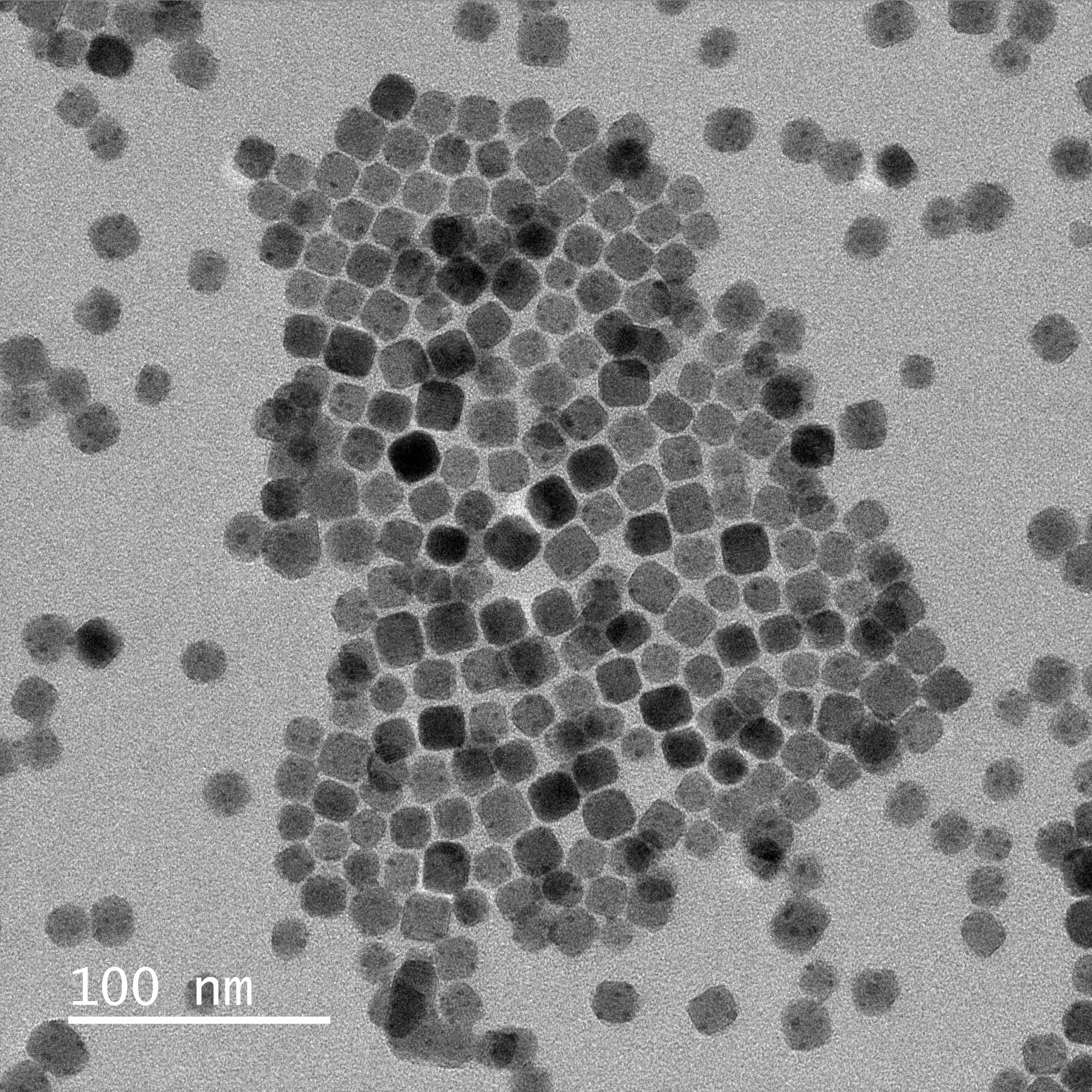}
        \caption{Representative TEM image of the PQDs.}
        \label{fig:subim1}
    \end{subfigure}
    \hfill % Adds horizontal space between the two subfigures
    \begin{subfigure}[t]{0.49\textwidth}
        \centering
        \includegraphics[width=\textwidth]{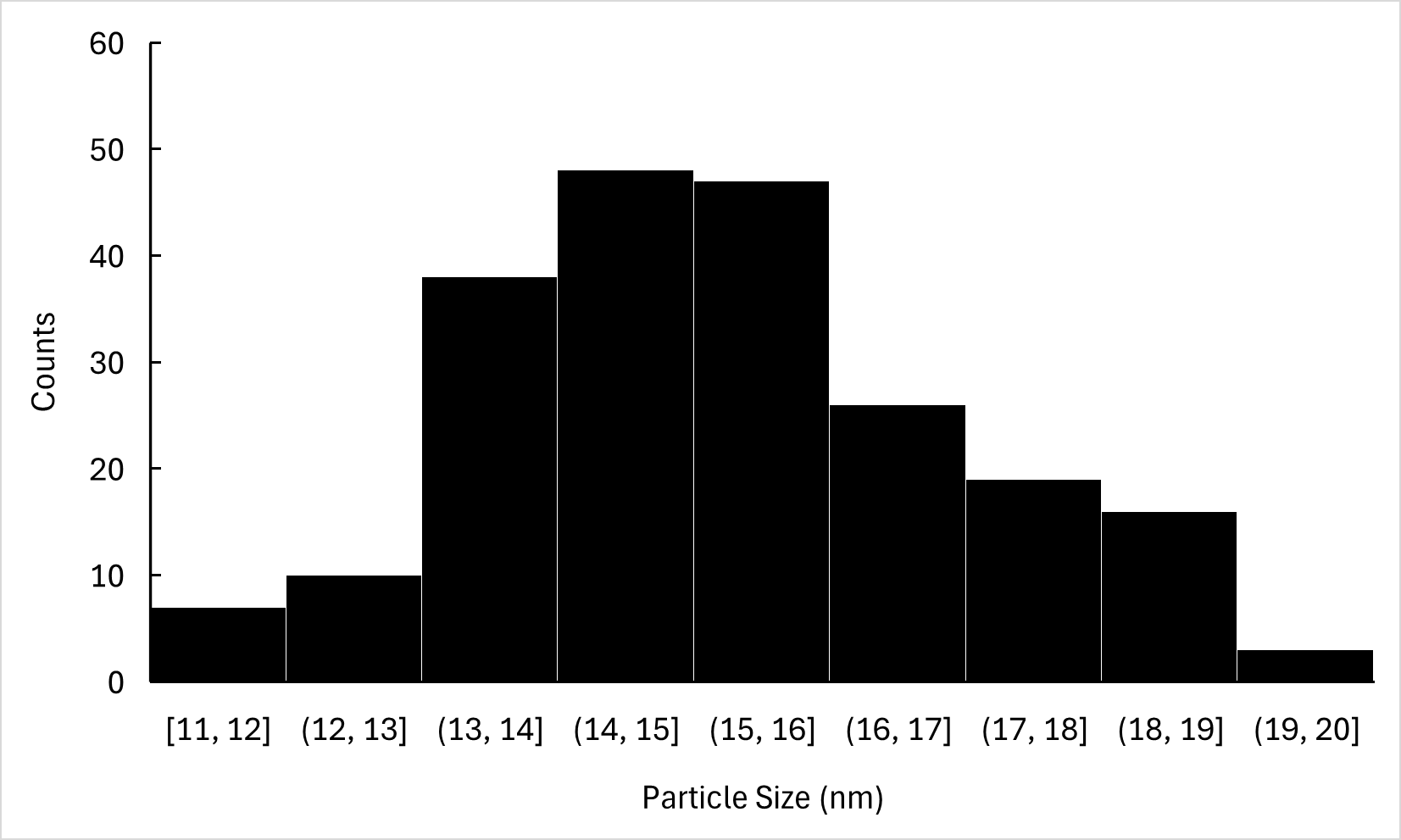}
        \caption{PQD size distribution based on the provided image.}
        \label{fig:subim2}
    \end{subfigure}
    \caption{Mean particle Edge length is 15.5 nm with a standard deviation of 1.8 nm. The distribution is based on a sample size of 216 measurements made using imageJ.}
    \label{fig:combined}    
\end{figure}

\begin{figure}
     \centering
     \begin{subfigure}{0.49\textwidth}
         \centering
         \includegraphics[width=\linewidth]{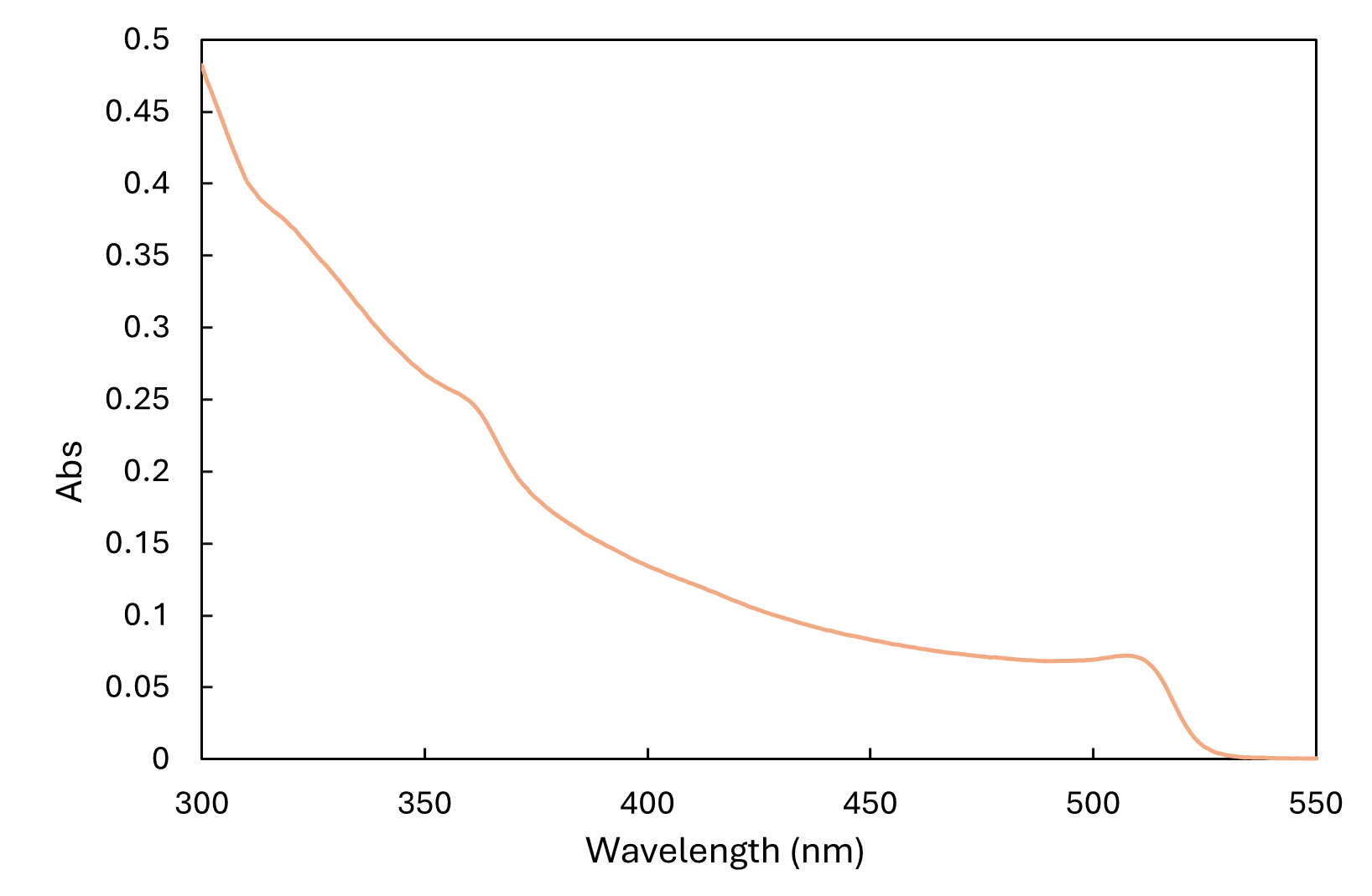}
         \caption{PQD ensemble absorbance spectrum}
         \label{fig:1a}
     \end{subfigure}
     \begin{subfigure}{0.49\textwidth}
         \centering
         \includegraphics[width=\linewidth]{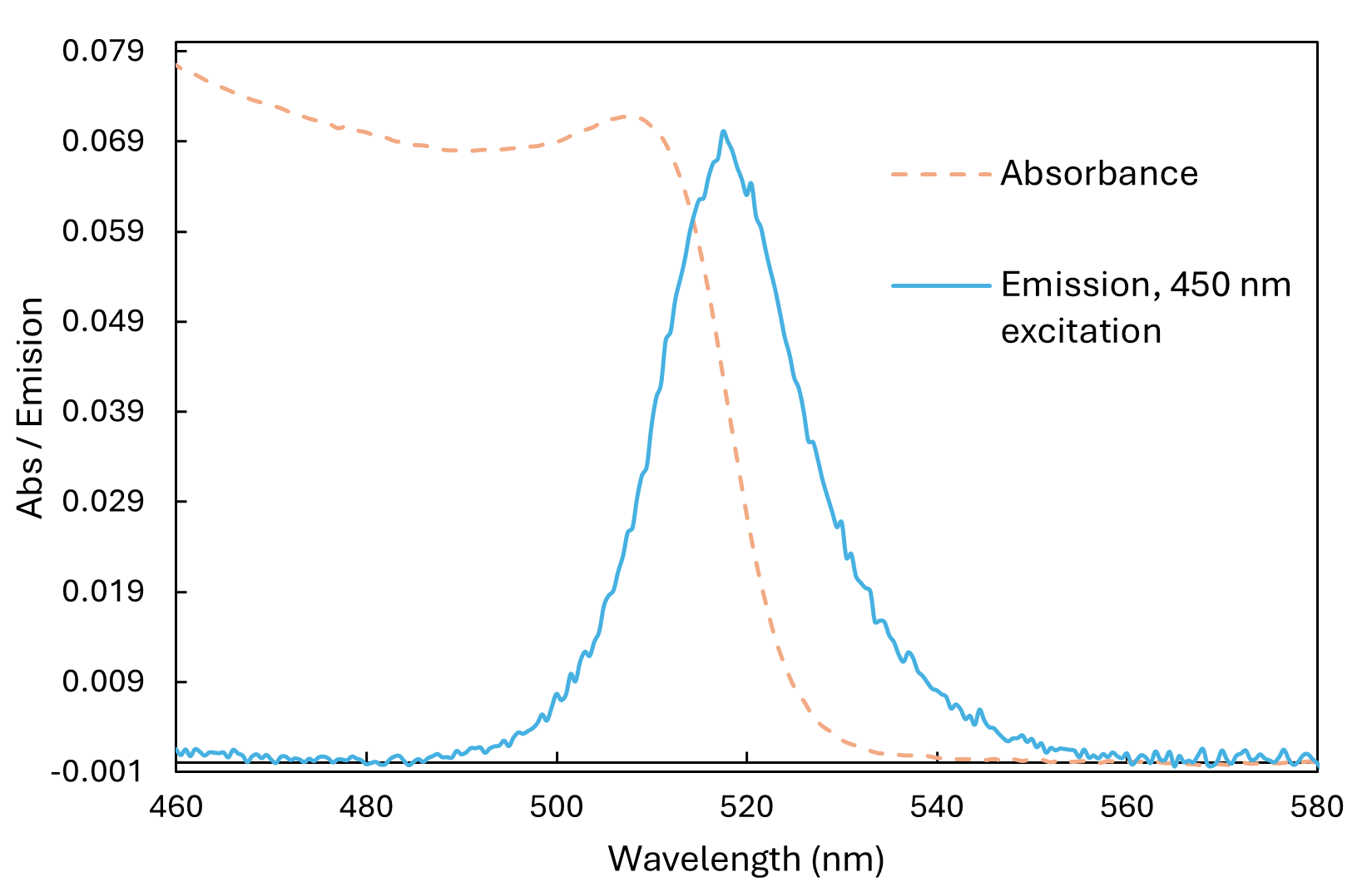}
         \caption{1st exciton and ensemble emission spectrum}
         \label{fig:1b}
     \end{subfigure}
     \caption{The peak of the first exciton is at 508 nm while the ensemble emission peak is at 517.5 nm. PQD emission has a FWHM of 21 nm and a PLQY of 58\% with 450 nm excitation. Both measurements were made in toluene.}
     \label{fig:1}
\end{figure}
\begin{figure}
    \centering
    \includegraphics[width=0.4\linewidth]{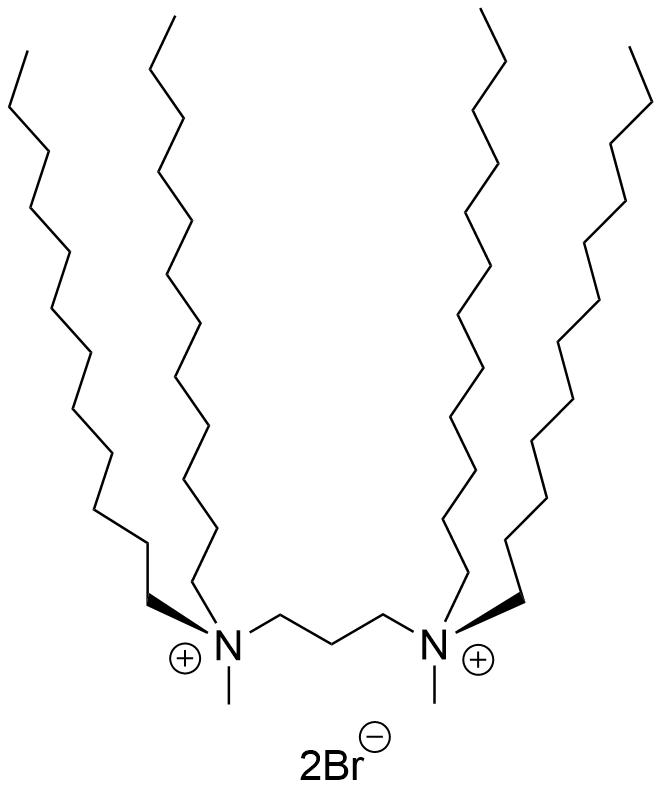}
    \caption{The structure of the C3-2C12 diquat bromide ligand used to synthesize the PQDs.}
    \label{fig:placeholder}
\end{figure}
\begin{figure}
    \centering
    \includegraphics[width=1.0\linewidth]{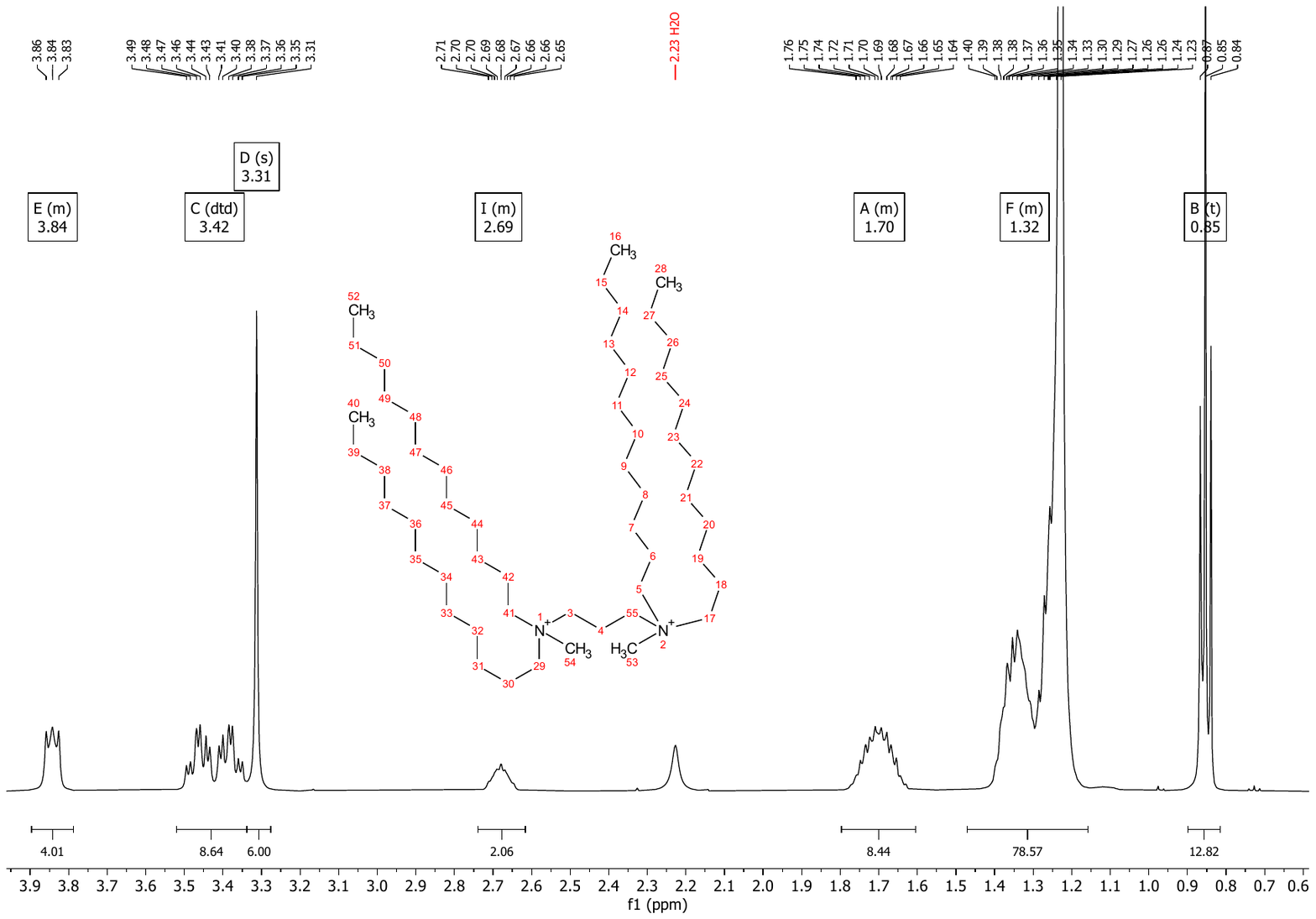}
    \caption{\textbf{C3-2C12} 1H NMR (500 MHz, CDCl3) $\delta$ 3.88 – 3.81 (m, 4H), 3.42 (dtd, J = 42.1, 12.8, 5.1 Hz, 8H), 3.31 (s, 6H), 2.75 – 2.59 (m, 2H), 1.78 – 1.63 (m, 8H), 1.44 – 1.13 (m, 74H), 0.85 (t, J = 6.9 Hz, 11H). Reported 1H NMR peaks align with the shifts reported by Ginterseder while peak integrals do not exactly align. This is likely due to impurities present in the starting material didodecylmethylamine ($>$85\%).}
    \label{fig:placeholder}
\end{figure}
\begin{figure}
    \centering
    \includegraphics[width=1\linewidth]{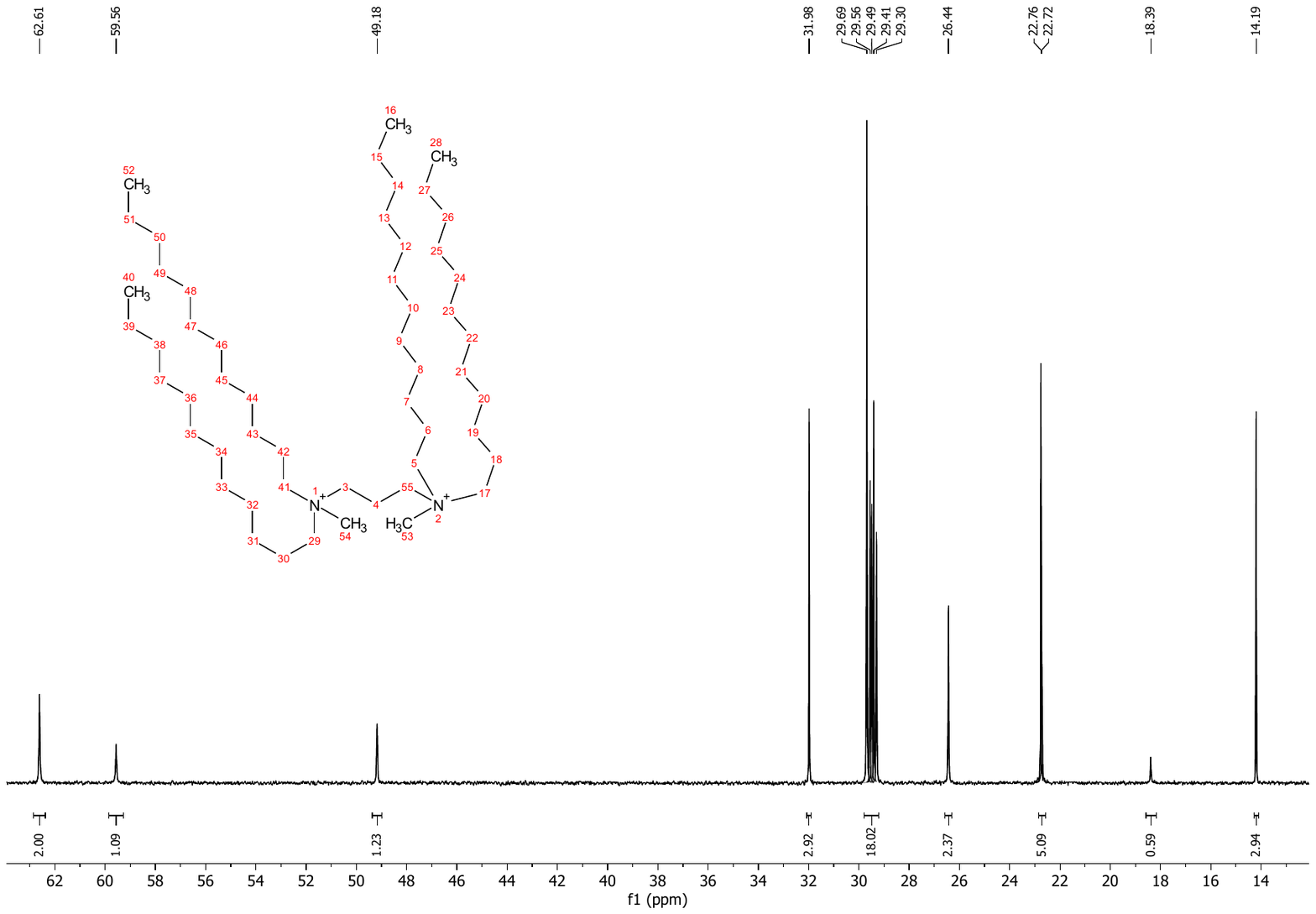}
    \caption{\textbf{C3-2C12} 13C NMR (126 MHz, CDCl3) $\delta$ 14.19, 18.39, 22.72, 22.76, 26.44, 29.30, 29.41, 29.49, 29.56, 29.69, 31.98, 49.18, 59.56, 62.61.}
    \label{fig:placeholder}
\end{figure}

\end{document}